\documentclass[manuscript]{aastex}

\newcommand{\hmpc}{\;h^{-1}{\rm Mpc}} 
\newcommand{\be}{\begin{equation}} 
\newcommand{\ee}{\end{equation}} 
 
\slugcomment{Accepted for publication in ApJL}

\shorttitle{Does the correlation length increase with depth?} 
\shortauthors{Mart{\'\i}nez et al.}
\begin{document} 
 
\title{Does the galaxy correlation length increase with the sample 
depth?} 
\author{Vicent J. Mart{\'\i}nez\altaffilmark{1}}
\affil{Observatori Astron{\`o}mic, 
Universitat de Val{\`e}ncia, E--46100 Burjassot, 
Val{\`e}ncia, Spain}
\email{vicent.martinez@uv.es}

\author{Bel{\'e}n L{\'o}pez--Mart{\'\i}\altaffilmark{2}} 
\affil{Th\"uringer Landessternwarte, D--07778 Tautenburg, Germany}
\email{belen@tls-tautenburg.de}

\and

\author{Mar{\'\i}a--Jes{\'u}s Pons--Border{\'\i}a\altaffilmark{3}} 
\affil{Departamento de Matem{\'a}tica Aplicada y Estad{\'\i}stica, Universidad
Polit{\'e}cnica de Cartagena, E--30203  Cartagena, Spain} 
\email{maria.pons@upct.es} 

\begin{abstract} 
We have analyzed the behavior of the correlation length, $r_0$, as a
function of the sample depth by extracting from the CfA2 redshift survey 
volume--limited samples out to increasing distances. For a fractal distribution, 
the value of $r_0$ would increase with the volume occupied by the sample. 
We find no linear increase for the CfA2 samples of the sort that would be
expected if the Universe preserved its small scale fractal character out to
the distances considered (60--100$\hmpc$). The results instead show 
a roughly constant value for $r_0$ as a function of the size of the sample, 
with small fluctuations due to local inhomogeneities and luminosity segregation.
Thus the fractal picture can safely be discarded.
\end{abstract} 
 
 
\section{Introduction} 

In recent years, the question of whether the Universe becomes
homogeneous on large scales or continues to exhibit fractal structure has been
much debated, with different analyses of several three-dimensional surveys of 
galaxies yielding opposite results \citep{dav97,pie97,guz97,cap98,sca98,joy99,
pee98,wu99,mar99}. The standard tool for analysis of clustering in galaxy 
catalogs is the two--point correlation function $\xi(r)$ \cite{pee80}, which is 
found to follow reasonably well a power--law $\xi(r)=(r_0/r)^{\gamma}$ at small
separations, $r < 10 \hmpc$.
The correlation length $r_0$ is, therefore, the scale
at which the correlation function $\xi(r)$ passes through unity. 
At a distance $r_0$ from an arbitrarily chosen galaxy, 
the number density of galaxies is on average twice the mean. 
A strong prediction of the fractal interpretation of 
galaxy clustering is an increase of the correlation length  
with the radius of the sample $R_s$ (Pietronero 1987; Guzzo 1997;
Sylos--Labini, Montuori, \& Pietronero 1998). 
\be 
r_0 = \left( {3- \gamma \over 6} \right) ^{1/\gamma} R_s \,. 
\label{pietro} 
\ee 
 
This kind of behavior is clearly observed in fractal point distributions. 
To illustrate it, we can use the fractal construction devised by 
Soneira \& Peebles (1978). 
The model is built as follows: Inside a sphere of radius $R$, place 
randomly $\eta$ spheres of radius $R/\lambda$ ($\lambda>1$).
Within each of these spheres place again $\eta$ spheres of radius
$R/\lambda^2$. The process is repeated $L$ times and
the last $\eta^L$ centers are considered galaxies.
A single clump is a simple fractal with a high degree of lacunarity, 
however Soneira \& Peebles 
(1978) used a superposition of a number of these clumps in order to mimic
the galaxy distribution provided by the Lick maps (Seldner et al. 1977). 
This construction has been recently taken 
by Peebles (1998) to show the extreme anisotropy of the 
distribution associated with this kind of fractal. 
 
We have built a model that avoids overlapping of the spheres at each step. 
The parameters chosen are $\eta=2, \lambda=2, R=60 \hmpc$ and $L=18$, which 
provides a dimensionality of $D_2 = 1$. In Figure~\ref{soneira} (upper 
left panel) we show the Aitoff projection of the model as seen by 
an observer situated at a point, close to the center of the simulation. 
The central left panel of Figure~\ref{soneira} shows the function 
$g(r)=1+\xi(r)$ for samples centered at this point and
radius ranging from 0.1 to 10 $\hmpc$. 
We can see that, as a consequence of the fractal character of this model, 
$g(r)$ is very nearly a power--law. The figure also shows that
the amplitude of $g(r)$ increases with the depth of the sample.
The correlation length, $r_0$, (shown in the bottom left panel of 
Figure~\ref{soneira}) increases with sample size as is expected for 
a fractal (Pietronero 1987). Fractals are intrinsically rather anisotropic 
patterns, but we can use a superposition of these fractal clumps to increase
the large--scale isotropy of the model as Soneira \& Peebles did
to mimic the Lick catalog. The price we have to pay is to lose
fractality at large scales, but since at small scales the fractal 
imprint of the model remains, the behavior of $r_0$ is still
the one expected for a fractal as we can see in the right panels
of Figure~\ref{soneira}. At the upper right panel we show a realization 
of a superposition of 125 clumps with parameters $\eta=2$,
$\lambda=1.76,  R=60 \hmpc$, and $L$ equals values taken from 
a Gaussian distribution with mean 6 and standard deviation 1. 
In this case overlapping is allowed. The two panels below show
$g(r)$ and $R_0(R_s)$ for this specific model.

For the galaxy distribution, the increase of $r_0$ with the depth of the sample 
was already noticed in the first redshift survey analyzed so far, the CfA1 
catalog (Einasto, Klypin, \& Saar 1986). The issue, however, 
is more complicated because problems such as 
local inhomogeneities, corrections for Galactic 
extinction and, most important, luminosity segregation play an important role 
in the behavior of $r_0$ with sample depth (Davis et al. 1988;
Mart{\'\i}nez et al. 1993; Benoist et al. 1996; Willmer, da Costa, \&
Pellegrini 1998, Beisbart \& Kerscher 2000).
 
The aim of this Letter is to see whether the correlation length increases with 
sample depth as expected in the fractal picture.
We have studied the best available redshift catalogue for this purpose: the
CfA2 survey (Geller \& Huchra 1989; Huchra, Vogeley \& Geller
1999). This sample is wide enough and deep enough to provide meaningful
results. The strategy will be to disentangle the possible volume effect from  
the segregation of luminosity.

\section{Description of the samples}

We have worked on several volume--limited samples of the CfA2
surveys, constructed in the following way:  
 
\emph{CfA2 north}. First, we have extracted
volume--limited samples as in Park et al. (1994).
Radial velocities were transformed into comoving coordinate distances 
by means of Mattig's formula (for $\Omega=1$):
 
\begin{equation} \label{rz}
r(z)= \frac{2c}{H_0} \left ( 1 - \frac{1}{\sqrt{1+z}} \right ),
\end{equation}
 
\noindent with $H_0= 100\,h$ km s$^{-1}$ Mpc$^{-1}$ and 
$c=299792.5$ km s$^{-1}$.
The angular region is $8^{\mbox{\rm h}} \leq \alpha \leq 16^{\mbox{\rm h}}$ 
(note that in Park et al. (1994) the upper
limit was $17^{\mbox{\rm h}}$) and $8.5^\circ \leq \delta \leq 44.5^\circ$.
The names of the volume--limited samples listed in Table 1
are of the form CfAn$d$, where $d$ is the rounded depth of the sample in
$\hmpc$. We list for each sample the number of galaxies $N_g$, the absolute
magnitude limit $M_{\ell}$, the depth $R_{\max}$,
the volume $V$, a typical length $R_s$ associated with the sample 
(the cubic root of the volume), and finally the correlation length $r_0$
which has been fitted to that sample. Note that 
$M_{\ell} = 15.5-5 \log[r(z)(1+z)]-25-Kz$, where 
$K=3$ is the appropriate
value of the $K$--correction for the $B$ filter and
the term $+5\log h$ has been
systematically omitted in the values of the magnitudes
throughout the paper.

In addition, we have drawn two more subsamples from CfAn101 having the same
absolute magnitude limit but different geometry. First, we have extracted 
the galaxies contained in the maximum parallelepiped completely
embedded within the boundaries of CfAn101
(see Figure~\ref{samples}, left panel).
The resulting sample, BOXn, has 432 objects and
a volume roughly 60\% of the parent sample.
Then we have calculated the  
depth that a sample with the same geometry as CfAn101 should reach in  
order to encompass the same volume as BOXn;
this distance  turned out to be  $84.79 h^{-1}$ Mpc.
A subsample of CfAn101 containing only galaxies
up to this distance is called CUTn (see Figure~\ref{samples}, central 
panel). Note that BOXn, CUTn and the volume--limited sample
CfAn85 have all the same volume. In fact, CUTn is just a subsample of CfAn85
containing its brightest galaxies ($M< -19.70$).

\emph{CfA2 south}. The samples extracted from the CfA2 south survey 
lie within the
angular region $22.5^{\mbox{\rm h}} \leq \alpha \leq 3^{\mbox{\rm h}}$ 
and $0^\circ \leq \delta \leq 40^\circ$.
In most of this region the Galactic extinction is low according to the maps
of Burstein and Heiles (1982). Two volume--limited samples have been drawn
from the survey, CfAs75 and CfAs59 (see Table 1) and from CfAs75
two subsamples have been extracted following
the same procedure as in the north, BOXs and CUTs.

\section{Results and discussion}

  For each sample the correlation function $\xi(r)$ has been calculated using  
the Rivolo estimator (Rivolo 1986),
\begin{equation} 
\xi (r) = \frac{V}{N^2} \sum_{i=1}^N \frac{n_i(r)}{V_i} -1,
\end{equation} 
where $n_i(r)$ is the number of galaxies lying 
at distance in the interval $[r-dr/2, r+dr/2]$ from galaxy $i$ and 
$V_i$ is the volume of the intersection  
of a shell with radii $r-dr/2$ and $r+dr/2$, centered at that galaxy, 
with the sample volume (logarithmic bins were used). 
Figure~\ref{xi} (left panel) shows
the results for all volume--limited samples extracted from the CfA north
survey. We have performed a weighted least--squares fit of $\xi(r)$   
in redshift space to a power-law $\xi(r)=(r_0/r)^{\gamma}$  
within the range $[3-10 \hmpc]$, where the fit is reasonably good.
This is an important point to bear in mind, since different definitions
of the clustering length are normally used in the literature and caution has 
to be exercised when intercomparing them (Peebles 1989). 
Poisson errors of the estimates of $\xi(r)$ have been used to weight the
fit. By this means we  
have calculated a value of $r_0$ for each sample together with an error
estimate. The results 
are reported in Table 1 and, for the volume--limited samples, in the right
panel of Figure~\ref{xi}. 
 
The most remarkable result is that the correlation length does not 
significantly change
with the depth of the sample. In any case, the linear increase
predicted for a fractal pattern is clearly ruled out.
For example, the volume of the sample CfAn101 is about three times
the volume of the sample CfAn70, while the correlation length, $r_0$,
of both samples is comparable. Note that the plateau around 
$r_0 \simeq 6.7 \hmpc$ observed in the
right panel of Figure~\ref{xi} includes also samples from the CfA2
south survey. In this case, we can stress that being the volume of 
the sample CfAs75 about twice the volume of the sample CfAs59, again
the corresponding $r_0$ values are practically the same.
It is interesting to note that when we have selected
the largest parallelepiped embedded within CfAn101, i.e. BOXn, the value of
the correlation length has increased although the volume has  
decreased nearly by half. Instead, the opposite behavior has been found 
in BOXs regarding its parent sample. Clearly the 
value of $r_0$ is more affected by the weight of local inhomogeneities
within the sample volume than by the change in the volume itself.
 
As we have explained, CUTn is a subsample of CfAn85.
Both samples lie within the same volume, but CUTn contains only galaxies
brighter than the absolute magnitude limit of CfAn101 $(-19.70)$.
The intrinsically brighter subsample, CUTn,
exhibits a larger value of the correlation length, $r_0 = 7.34 \pm 0.78 \hmpc$, 
than the whole sample, CfAn85, for which $r_0 = 6.43 \pm 0.30 \hmpc$. 
This effect is a fingerprint of the luminosity segregation and it
is also observed in the CfA2 south samples, when we compare the value of
$r_0$ corresponding to the sample CUTs, $r_0 = 7.54 \pm 0.63 \hmpc$, 
with the value
corresponding to the sample CfAs59, $r_0 = 6.34 \pm 0.28 \hmpc$.
The dependence of the
correlation length on galaxy luminosity found by us is consistent with
the results reported by Park et al. (1994), analyzing the amplitude
of the power spectrum for the same survey. 

Because of the effect of random peculiar motions, the correlation length 
in real space should be smaller by approximately 1 $\hmpc$ than
the correlation length in redshift space (Peebles 1989, 
Loveday et al. 1995). Values of the redshift space 
correlation length near 6.5 $\hmpc$ would therefore correspond to values 
$\sim 5.5 \hmpc$ in real space (Davis \& Peebles 1983). 

In sum, we conclude that the correlation length calculated on
volume--limited samples extracted from the CfA2 redshift survey
is a rather stable quantity. In any case, the linear increase
with radius of the sample predicted for a fractal is not
observed. Our conclusions are reinforced by the fact that
the correlation length in redshift space is still about 6 $\hmpc$ 
in redshift surveys that go much deeper than CfA2 (though they are sparser
or narrower), like the Stromlo-APM, Las Campanas, and ESP redshift surveys
 (Loveday et al. 1995; Tucker et al. 1997; Guzzo et al. 2000).
 On still larger scales,  Gladders \& Yee (2000) have shown that
the correlation length for luminous early-type galaxies does not
change from $z=0$ to $z=1$, and Carlberg et al. (2000) find that the
correlation amplitude declines slightly with redshift over the range
0.1 to 0.6 in the CNOC2 sample of very luminous galaxies.

\acknowledgements 
We are grateful to J.P. Huchra for providing us with the CfA2 data.
We thank S. Paredes and J.J. Nu\~no for suggestions and very especially
V. Trimble for a careful reading of the manuscript. This work was 
supported by the Spanish Ministerio de Ciencia y Tecnolog\'{\i}a project 
AYA2000-2045.

\clearpage

        \begin{deluxetable}{lrlrrcc}
        \tablecolumns{7} 
        \tablecaption{Characteristics of the subsamples of the CfA2 catalogue}
        \label{tab1} 
        \tablewidth{0pt} 
        \tablehead{ 
        \colhead{} & \colhead{} &\colhead{} &\colhead{$R_{\max}$} &\colhead{$V$} 
        & \colhead{$R_s$}& \colhead{$r_0$} \\ 
        \colhead{Sample} & \colhead{$N_{\rm g}$} & \colhead{$M_{\ell}$} &  
        \colhead{($\!\hmpc$)}& 
        \colhead{($\!\hmpc$)$^3$}& \colhead{($\!\hmpc$)}& \colhead{($\!\hmpc$)} 
        } 
       \startdata 
       \multicolumn{7}{c}{CfA2 north} \\
        \tableline         
        CfAn101 & 905 & -19.70 & 101.11 & 399160.64 & 73.63 & 6.88 $\pm$ 0.49 \\
        \noalign{\smallskip} 
        BOXn & 432 & -19.70 & 101.11 & 235390.79 & 61.74 & 8.04 $\pm$ 0.84 \\
         \noalign{\smallskip} 
       CUTn & 444 & -19.70 & 84.79 & 235390.79 & 61.74 & 7.34 $\pm$ 0.78 \\
      \noalign{\smallskip} 
        CfAn92 & 1074 & -19.50 & 92.26 & 303242.65 & 67.18 & 7.19 $\pm$ 0.39 \\
        \noalign{\smallskip} 
        CfAn85 & 1134 & -19.29 & 84.79 & 235390.79 & 61.74 & 6.43 $\pm$ 0.30 \\
        \noalign{\smallskip} 
        CfAn78 & 1159 & -19.10 & 78.19 & 184608.80 & 56.94 & 6.88 $\pm$ 0.32 \\
        \noalign{\smallskip} 
        CfAn70 & 1113 & -18.85 & 70.10 & 133041.53 & 51.05 & 6.65 $\pm$ 0.29 \\
        \noalign{\smallskip} 
        CfAn60 & 736 & -18.49 & 59.95 & 83167.33 & 43.65 & 4.21 $\pm$ 0.31 \\
 \noalign{\smallskip} 
        \tableline 
        \multicolumn{7}{c}{CfA2 south} \\
        \tableline         
        CfAs75 & 722 & -19.0 & 74.79 & 105598.94 & 47.27 & 6.61 $\pm$ 0.39 \\
        \noalign{\smallskip} 
        BOXs & 435 & -19.0 & 74.79 & 51705.30 & 37.25 & 5.70 $\pm$ 0.44 \\
         \noalign{\smallskip} 
        CUTs & 427 & -19.0 & 58.95 & 51705.30 & 37.25 & 7.54 $\pm$ 0.63 \\
         \noalign{\smallskip} 
        CfAs59 & 844 & -18.46 & 58.95 & 51705.30 & 37.25 & 6.34 $\pm$ 0.28 \\
         
        \enddata 
        \end{deluxetable}

\clearpage
\begin{figure}
\caption{Left panels. \emph{Upper panel}: Equal--area
projection of the Soneira-Peebles model as seen from an object
near the center of the model. We have considered only the points
lying within a spherical volume of radius 1/3 of the radius of the first
sphere. \emph{Central panel}: The function $g(r)$ calculated
on eight spherical samples of increasing radius in our fractal model.
The values of the radii are shown in the legend.
The reference line shows the expected slope for this fractal pattern.
\emph{Bottom panel}: Correlation length $r_0$ versus sample
radius $R_s$. In the right panels, 
the same figures are displayed for a model
having more large-scale isotropy, built by means of a superposition of 125
single fractal clumps similar to the one shown on the top left panel, 
although having different parameters (see text).} 
\label{soneira} 
\end{figure} 

\begin{figure}
\caption{\emph{Left panel}:
The sample CfAn101 containing the parallelepiped which is the
boundary of
the subsample BOXn (galaxies within BOXn have been highlighted in red).
\emph{Central panel}: The same sample, but now the highlighted
subsample is CUTn. \emph{Right panel}:
The sample CfAs75 containing the parallelepiped which is the
boundary of
the subsample BOXs (galaxies within BOXs have been highlighted in red).}
\label{samples} 
\end{figure}

\begin{figure}
\caption{ \emph{Left panel}: The correlation function for all the
volume--limited samples drawn from the CfA2 north survey (CfAn$d$ in 
Table 1).
All of them are power
laws with about the same amplitude, except for the closest sample CfAn60, 
which has a significantly smaller amplitude. Vertical dashed lines
show the range where we have fitted a power--law to $\xi(r)$ to obtain
the value of $r_0$. 
\emph{Right panel}: The value of 
the correlation length for all the
volume--limited samples (CfAn$d$ and CfAs$d$ in Table 1). Again, apart
from the small value for CfAn60, the rest of the samples yield
values of $r_0$ lying within the range $[6.4 - 7.2 \hmpc]$.}
\label{xi}
\end{figure} 
    
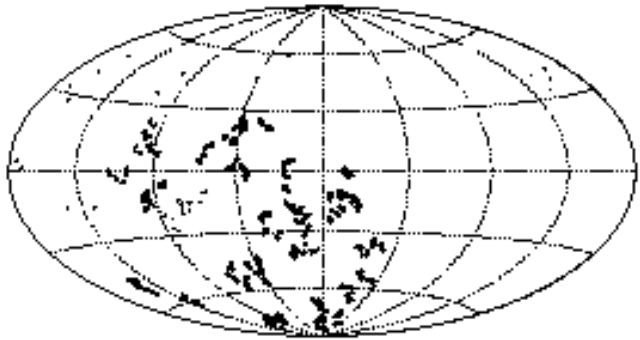
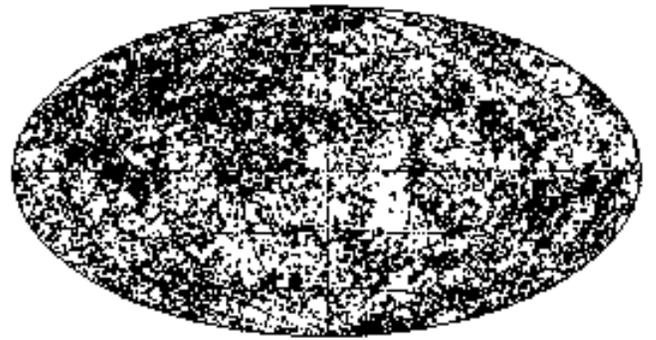
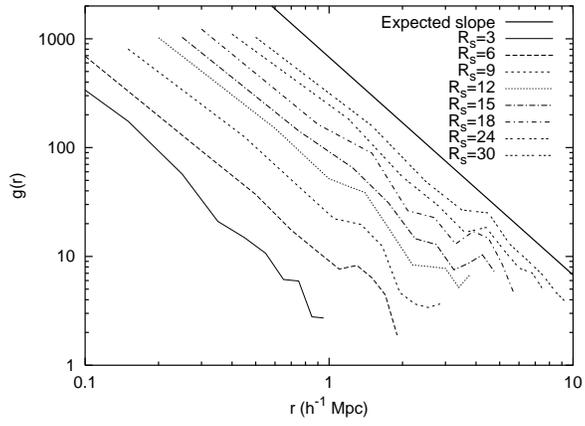
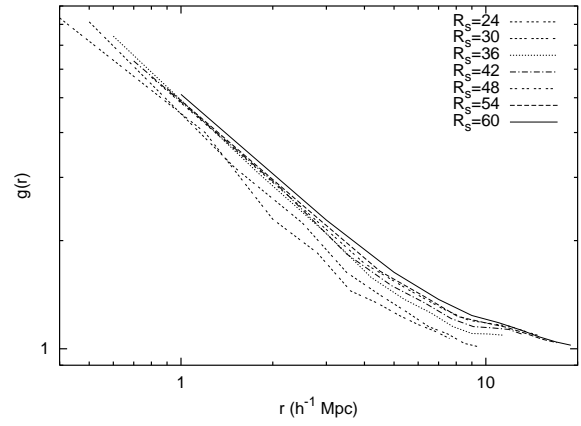
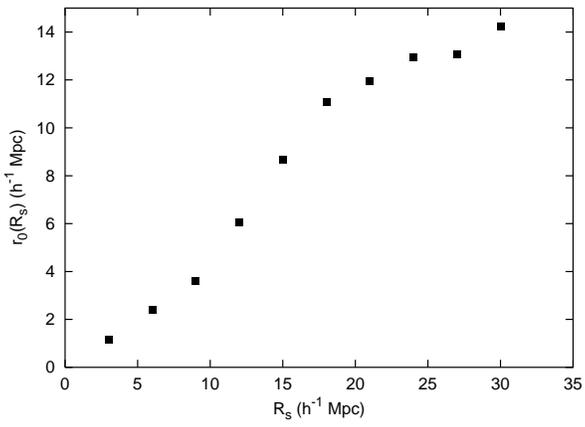
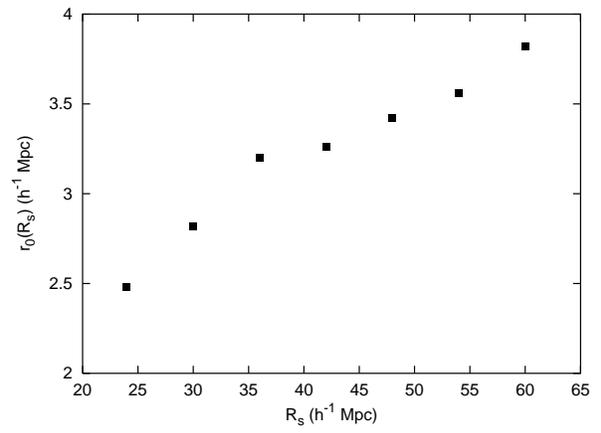
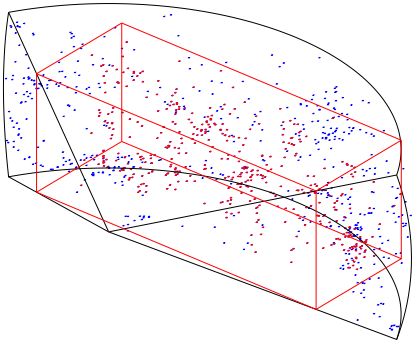
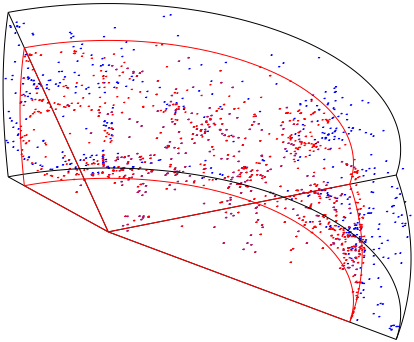
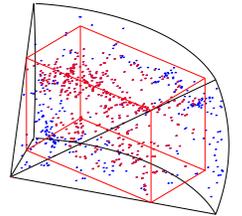
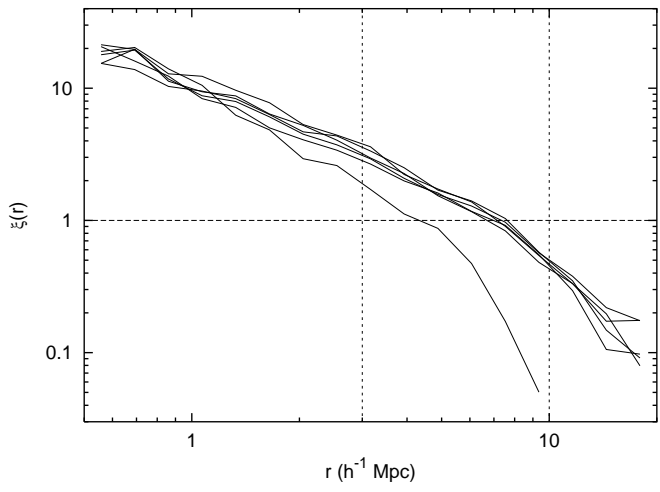
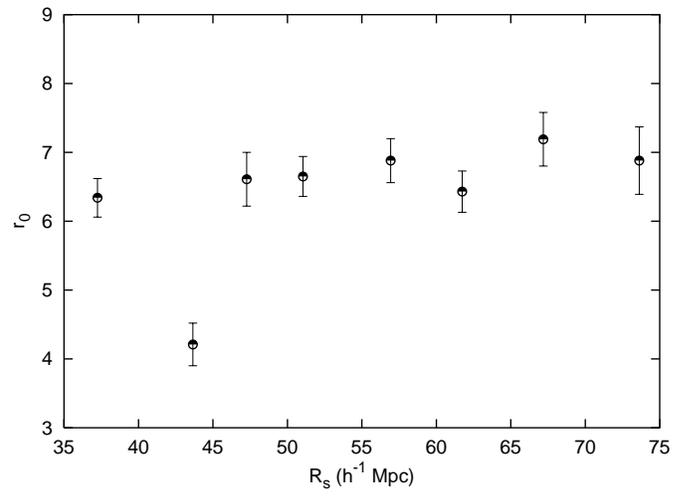
\end{document}